\immediate\write18{makeindex \jobname.nlo -s nomencl.ist -o \jobname.nls}

\documentclass[conference]{IEEEtran}

\usepackage[table,xcdraw]{xcolor}

\usepackage{booktabs}
\newcommand{\ra}[1]{\renewcommand{\arraystretch}{#1}}

\usepackage{multirow}
\usepackage{color}

\usepackage{ifpdf}
\ifpdf
\fi

\usepackage{cite}

\ifCLASSINFOpdf
\usepackage{graphicx}
\usepackage{float}
\ifCLASSOPTIONcompsoc
\usepackage[caption=false, font=normalsize, labelfont=sf, textfont=sf]{subfig}
\else
\usepackage[caption=false, font=footnotesize]{subfig}
\fi

\usepackage{amsmath}
\usepackage{amssymb}
\usepackage{fixltx2e}

\usepackage{microtype} 
\usepackage[table,xcdraw]{xcolor} 

\usepackage{nomencl}
\makenomenclature

\usepackage{epstopdf}

\usepackage{soul} 

\usepackage{epstopdf}
\usepackage{scalerel}
\usepackage{tikz}
\usetikzlibrary{svg.path}

\definecolor{orcidlogocol}{HTML}{A6CE39}
\tikzset{
	orcidlogo/.pic={
		\fill[orcidlogocol] svg{M256,128c0,70.7-57.3,128-128,128C57.3,256,0,198.7,0,128C0,57.3,57.3,0,128,0C198.7,0,256,57.3,256,128z};
		\fill[white] svg{M86.3,186.2H70.9V79.1h15.4v48.4V186.2z}
		svg{M108.9,79.1h41.6c39.6,0,57,28.3,57,53.6c0,27.5-21.5,53.6-56.8,53.6h-41.8V79.1z M124.3,172.4h24.5c34.9,0,42.9-26.5,42.9-39.7c0-21.5-13.7-39.7-43.7-39.7h-23.7V172.4z}
		svg{M88.7,56.8c0,5.5-4.5,10.1-10.1,10.1c-5.6,0-10.1-4.6-10.1-10.1c0-5.6,4.5-10.1,10.1-10.1C84.2,46.7,88.7,51.3,88.7,56.8z};
	}
}

\newcommand\orcidicon[1]{\href{https://orcid.org/#1}{\mbox{\scalerel*{
				\begin{tikzpicture}[yscale=-1,transform shape]
					\pic{orcidlogo};
				\end{tikzpicture}
			}{|}}}}
		
\usepackage{hyperref} 
\hypersetup{hidelinks}

\makeatletter
\def\@fnsymbol#1{\ensuremath{\ifcase#1\or \dagger\or \ddagger\or
		\mathsection\or \mathparagraph\or \|\or **\or \dagger\dagger
		\or \ddagger\ddagger \else\@ctrerr\fi}}
\makeatother

\IEEEoverridecommandlockouts
\begin{document}

\title{A Guideline for Silicon Carbide MOSFET Thermal Characterization based on Source-Drain Voltage}


\author{\IEEEauthorblockN{
	Yi~Zhang$^\dagger$
	\orcidicon{0000-0003-0248-7644}\,,
	Yichi~Zhang$^\dagger$,
	Zhiliang~Xu$^\ddagger$,
	Zhongxu~Wang$^\mathsection$, 
	Hon~Wong$^*$, 
	Zhebie~Lu$^\dagger$, 
	and~Antonio~Caruso$^*$}
	\IEEEauthorblockA{$^\dagger$AAU Energy, Aalborg University, Aalborg, Denmark\\
	$^\ddagger$Southwest Jiaotong University, Chengdu, China\\
	$^\mathsection$Nexperia UK LTD, United Kingdom\\
	$^*$Siemens Digital Industries Software, United States\\
	Email: yiz@ieee.org}
	
}

\markboth{IEEE LATEX, Vol. X, No. X, Month, Year}%
{Shell \MakeLowercase{\textit{et al.}}: Bare Demo of IEEEtran.cls for IEEE Journals}

\maketitle

\begin{abstract}
Thermal transient measurement based on source-drain voltage is a standard method to characterize thermal properties of silicon semiconductors but is doubtful to be directly applied to silicon carbide (SiC) devices. To evaluate its feasibility and limitations, this paper conducts a comprehensive investigation into its accuracy, resolution, and stability towards yielding the structure information of SiC MOSFET using the source-drain voltage as the temperature sensitive electrical parameter. The whole characterization process involves two main procedures and associated key testing parameters, such as gate voltages, sensing and heating currents, etc. Their impacts on both the static and dynamic performances are also investigated with the aim of providing a guideline for conducting a reproducible thermal transient measurement for SiC MOSFETs.
\end{abstract}


\IEEEpeerreviewmaketitle


\section{Introduction}

Thermal transient measurement is a widely accepted method to characterize thermal properties of silicon (Si) power semiconductor devices, and it has been recognized in several international standards, such as JEDEC JESD 51-1\cite{jesd511intergrated} and JEDEC 51-14\cite{JEDEC51-14}. In the past two decades, this method has been successfully applied to different applications, such as generating $RC$ thermal models for electro-thermal simulation, package defects inspection\cite{gao2020two}, junction-to-case thermal resistance measurement\cite{schweitzer2008transient}. However, directly applying this approach to SiC MOSFETs is still challenging\cite{kestler2018junction,yangfei2019design}.

\begin{figure*}[!t]
	\centering
	\includegraphics[width=170mm]{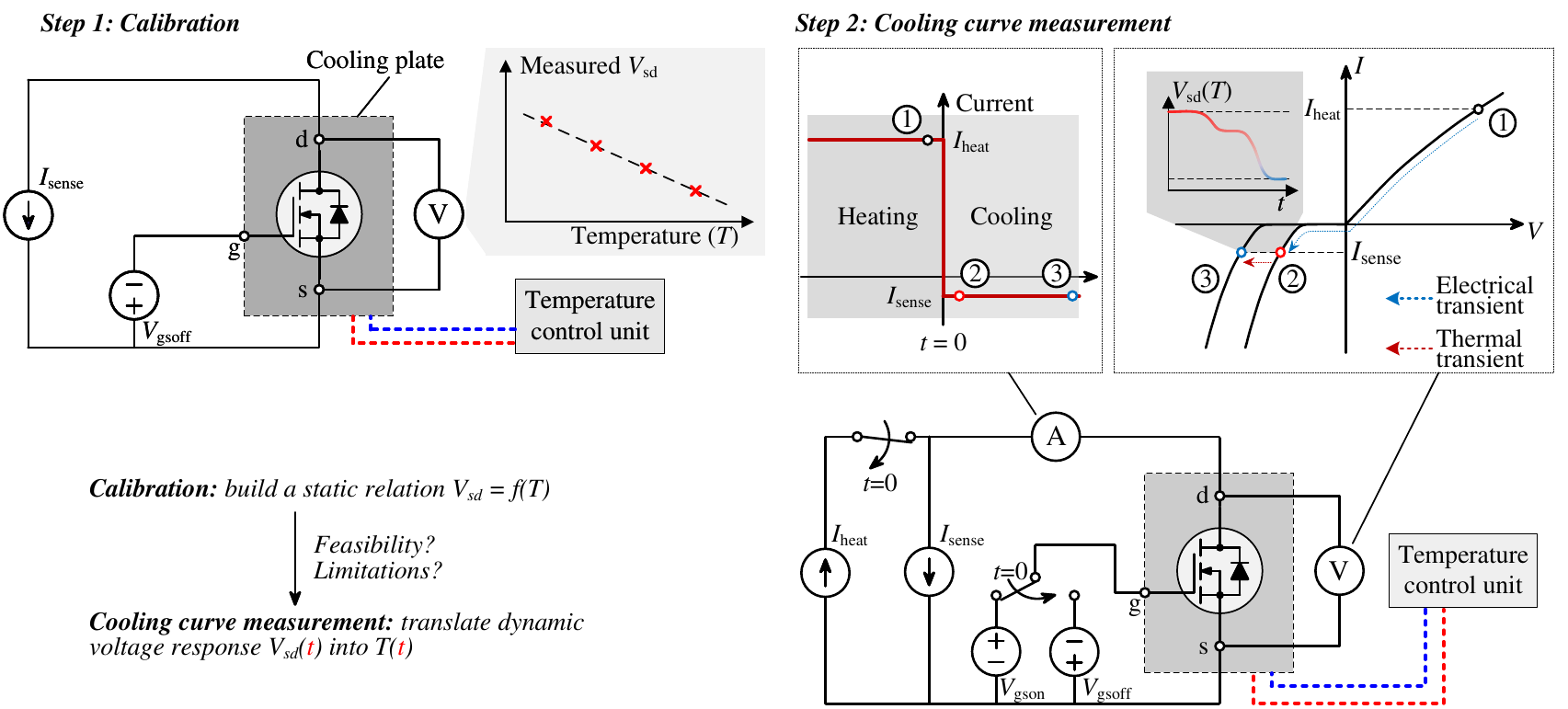}
	\caption{Circuit diagram for temperature calibration (step 1) and cooling curve measurement (step 2) for SiC MOSFETs. \emph{Calibration}: the device is mounted on a temperature-controlled cooling plate. The $V_\text{sd}$ under a small sensing current $I_\text{sense}$ is measured as a function of the temperature. A gate turn-off voltage $V_\text{gsoff}$ is used to shut MOS channel off completely. \emph{Cooling curve measurement}: a large heating current $I_\text{heat}$ heat up the device to a thermal equilibrium $\textcircled{1}$. Subsequently, the current is dramatically reduced to $I_\text{sense}$ at $\textcircled{2}$. The transient of $\textcircled{1} \rightarrow \textcircled{2}$ is the unwanted electrical transient, and the time-resolved response of $\textcircled{2} \rightarrow \textcircled{3}$ is the measured cooling curve.}
	\label{fig:calibration}
\end{figure*}

Compared to Si-based devices (e.g., Si IGBTs)\cite{bhatnagar1993comparison}, SiC MOSFETs do not have a pn junction in the forward direction and have low on-state resistance, which imposes challenges to measure transient thermal response by the channel voltage\cite{gonzalez2017impact}. Meanwhile, trapped charge carriers in the gate region may cause second-level electrical disturbances\cite{funaki2016difficulties}, which inevitably affect the extraction of thermal transient out of the coupled electrical disturbance which is as short as microseconds. In the state-of-the-art, the source-drain voltage is one of the most used temperature sensitivity electrical parameters (TSEP) for SiC MOSFETs. As shown in Fig.~\ref{fig:calibration}, the characterization consists of two major procedures, namely calibration and cooling curve measurement. Selecting improper testing parameters, e.g., the gate voltage, sensing or heating currents, etc., may lead to misleading results.

\begin{figure}[!t]
	\centering
	\includegraphics[width=75mm]{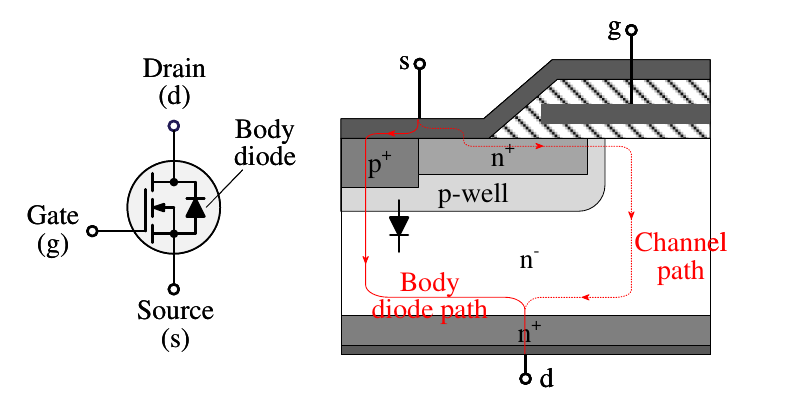}
	\caption{Structure of a SiC MOSFET. The pn junction of the body diode is used as the TSEP. A parallel current path exists in the MOS channel which may affect temperature measurement.}
	\label{fig:sic_mosfet}
\end{figure}

Therefore, this paper comprehensively investigates the thermal transient measurement approach of SiC MOSFETs using $V_\text{sd}$ as the TSEP and focuses on how to obtain the correct thermal structural information without considering the long-term aging effects that are out of the scope of this paper. The contributions are two folds: 1) the impact of key parameters including the gate turn-off voltage and the sensing current on the \emph{calibration} are evaluated, and 2) the impact of different parameters on the \emph{cooling curve measurement} is also investigated. Last but not least, a guideline is summarized to help perform a reproducible thermal transient measurement of SiC MOSFETs.

\section{Thermal Transient Measurement based on the Source-Drain Voltage of SiC MOSFETs}

Fig.~\ref{fig:calibration} illustrates how to obtain a time-resolved transient thermal response and the structural information of the device under test through two major steps, namely, calibration and cooling curve measurement.

The calibration is used to determine the relationship between the TSEP and the device temperature, which is controlled by an external system (e.g, an oven, a dielectric bath, or a temperature-controlled cooling plate). Similar to the $V_\text{ce}$ of IGBTs \cite{zeng2018difference}, MOSFET body diode pn-junction voltage $V_\text{sd}$ shows a linear temperature dependence given a small sensing current going through the device. By measuring $V_\text{sd}$ under this small constant sensing current and various temperatures, the relation of $V_\text{sd}=f(T)$ can be calibrated. Noted that a low enough negative gate voltage has to be applied to completely shut the MOSFET channel off during this process (see Fig.~\ref{fig:sic_mosfet}).

As the second step, cooling curve measurement is carried out based on two current levels: one is the heating current ($I_\text{heat}$) to heat the device up, and the other is the sensing current for temperature monitoring with a negligible self-heating impact, as shown in Fig.~\ref{fig:calibration} (Step 2). Once $V_\text{sd}$ is available, the inversely calibrated result $T=f^{-1}(V_\text{sd})$ in step 1 converts the measured voltage into the temperature transient. However, the temperature calibration is developed based on static conditions, while the cooling curve is derived from dynamic voltage responses. The compatibility of the two steps for Si IGBTs and MOSFETs has been validated by enormous studies, but it is not that clear for SiC MOSFETs. For instance, reference\cite{funaki2016difficulties} pointed out that the electrical transient is much longer for SiC MOSFETs. The complex coupling of electrical and thermal transients may impair the feasibility of the assumption used in Si devices.

To fully understand the limitations of applying the Si-friendly approach to SiC MOSFETs, this paper will use the following two sections to investigate the impacts of different testing parameters along the two procedures. Both static and dynamic behaviors will be studied comprehensively in order to guide a reproducible testing for SiC MOSFETs.

\begin{figure*}[!t]
	\centering
	\includegraphics[width=165mm]{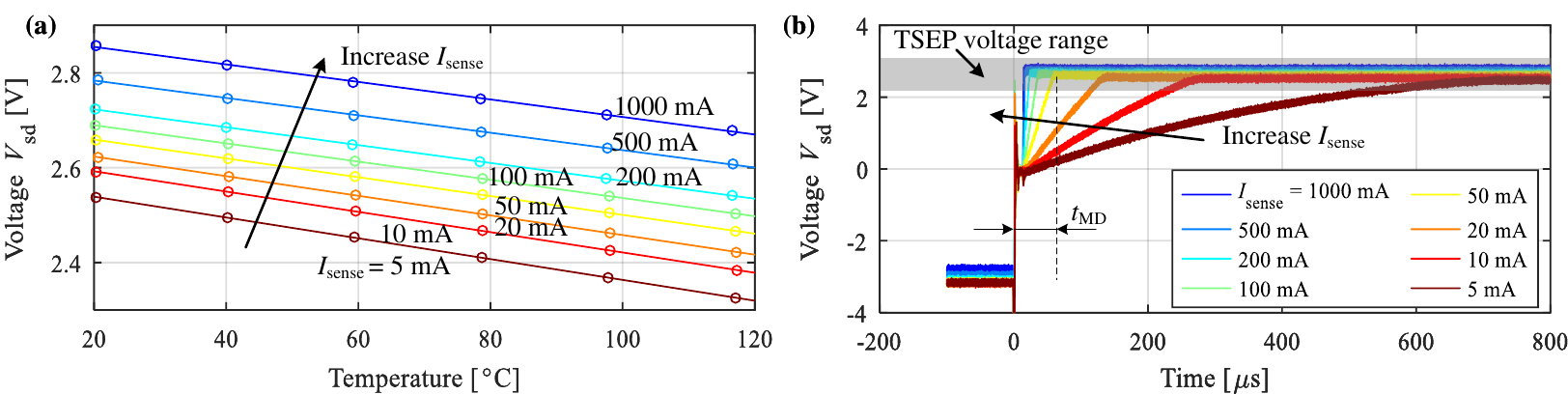}
	\caption{Calibration curves for multiple sensing currents with $V_\text{gsoff}=-6$~V: (a) static states and (b) dynamic states.}
	\label{fig:sense_current}
\end{figure*}
\begin{table*}[]
	\centering
	\ra{1.3}
	\caption{Calibration results under different sensing currents and gate turn-off voltages.}
	\label{table:Isense1}
	\scriptsize
	\begin{tabular}{ccccc|ccccc}
		\hline
		\multicolumn{5}{c|}{$V_\text{gsoff}=-6$ V (related to $\mathsection$\ref{sec:3a}, \ref{sec:dynamic_Isense})}                                                                                                                                                                                                                                                   & \multicolumn{5}{c}{$I_\text{sense}=100$ mA (related to $\mathsection$\ref{sec:static_gate_voltage}, \ref{sec:3d})}                                                                                                                                                                                                                                                 \\
		\begin{tabular}[c]{@{}c@{}}$I_\text{sense}$\\ {[}mA{]}\end{tabular} & Linearity & \begin{tabular}[c]{@{}c@{}}Resolution\\ {[}mV/K{]}\end{tabular} & \begin{tabular}[c]{@{}c@{}}Self dissipation\\ ratio\end{tabular} & \begin{tabular}[c]{@{}c@{}}$t_\text{MD}$\\ {[}$\mu$s{]}\end{tabular} & \begin{tabular}[c]{@{}c@{}}$V_\text{gsoff}$\\ {[}V{]}\end{tabular} & Linearity & \begin{tabular}[c]{@{}c@{}}Resolution\\ {[}mV/K{]}\end{tabular} & \begin{tabular}[c]{@{}c@{}}Self dissipation\\ ratio\end{tabular} & \begin{tabular}[c]{@{}c@{}}$t_\text{MD}$\\ {[}$\mu$s{]}\end{tabular} \\ \hline
		5                                                                   & 0.999948  & 2.192245                                                         & 0.022\%                                                          & 663                                                                  & 0                                                                  & 0.998424  & 2.987167                                                         & 0.281\%                                                          & 35                                                                   \\
		10                                                                  & 0.999945  & 2.139215                                                         & 0.045\%                                                          & 268                                                                  & -1                                                                 & 0.997931  & 2.576568                                                         & 0.342\%                                                          & 39                                                                   \\
		20                                                                  & 0.999954  & 2.068415                                                         & 0.091\%                                                          & 139                                                                  & -2                                                                 & 0.998039  & 2.145706                                                         & 0.399\%                                                          & 40                                                                   \\
		50                                                                  & 0.999968  & 1.983309                                                         & 0.230\%                                                          & 62                                                                   & -3                                                                 & 0.999339  & 1.704021                                                         & 0.445\%                                                          & 41                                                                   \\
		100                                                                 & 0.999995  & 1.918120                                                         & 0.467\%                                                          & 42                                                                   & -4                                                                 & 0.999992  & 1.873358                                                         & 0.465\%                                                          & 42                                                                   \\
		200                                                                 & 0.999940  & 1.884273                                                         & 0.943\%                                                          & 28                                                                   & -5                                                                 & 0.999992  & 1.921063                                                         & 0.468\%                                                          & 42                                                                   \\
		500                                                                 & 0.999678  & 1.835086                                                         & 2.411\%                                                          & 20                                                                   & -6                                                                 & 0.999995  & 1.918120                                                         & 0.467\%                                                          & 42                                                                   \\
		1000                                                                & 0.999415  & 1.847177                                                         & 4.947\%                                                          & 19                                                                   & -7                                                                 & 0.999996  & 1.923854                                                         & 0.467\%                                                          & 42                                                                   \\
		&           &                                                                  &                                                                  &                                                                      & -8                                                                 & 0.999995  & 1.921815                                                         & 0.467\%                                                          & 42                                                                   \\ \hline
	\end{tabular}
\end{table*}

\section{Impact of Calibration Parameters}\label{sec:calibration}
As shown in Fig.~\ref{fig:calibration} (step 1), the calibration of SiC MOSFETs based on $V_\text{sd}$ relies on two adjustable parameters, namely the gate turn-off voltage $V_\text{gsoff}$ and sensing current $I_\text{sense}$ to achieve good linearity, resolution, and a negligible small power dissipation in the static state. Meanwhile, short electrical disturbance is preferable in the dynamic state, which will also be covered in this section.

\subsection{Static Impact of Sensing Current Density}\label{sec:3a}
\subsubsection{Linearity}
The physical background of utilizing the pn-junction voltage drop $V_\text{pn}$ as the TSEP is its linear temperature dependence, which is given by
\begin{equation}\label{eq:pn_junc}
	{V_\text{pn}} = \frac{{{E_g}}}{q} - \frac{{{k_b} \cdot T}}{q}\ln \left( {\frac{K}{j_\text{sense}}} \right)
\end{equation}
where $E_g$ is the band gap, $q$ is the elementary charge, $k_b$ is the Boltzmann constant, $K$ is a device specific factor\cite{Lutz2011Semi_book}. All these parameters are independent of temperature or only have a weak temperature dependence. Given a constant sensing current density $j_\text{sense}$, the voltage drop $V_\text{pn}$ is linear with temperature $T$ according to (\ref{eq:pn_junc}). However, the measured $V_\text{sd}$ not only depends on the pn-junction voltage drop but also the resistance of the drift region, the contact and metallization as shown in Fig.~\ref{fig:sic_mosfet}, whose voltage contribution can be a dominant part of the measured $V_\text{sd}$ given a high sensing current. On the other hand, due to the negative temperature coefficient of the body diode, $V_\text{pn}$ can become very small given a high temperatures and a low current density and may lead to a nonlinear temperature correlation. Thus, a proper sensing current must be carefully justified prior to temperature measurements.

Fig.~\ref{fig:sense_current}(a) shows the calibration results for different sensing currents. Due to the three times higher band gap $E_g$ of SiC MOSFET than Si devices, the curves show higher intercepts even using a low sense current. Meanwhile, $V_\text{sd}$ becomes larger with the increase of the sensing current and is linear with temperature when the pn-junction voltage drop dominates the measurement loop (\ref{eq:pn_junc}). Thus, the linearity between $V_\text{sd}$ and the temperature can be used as a measure to justify whether the sensing current is properly selected. In this paper, Pearson correlation $\rho_\text{linear}$ is used as a justification based on its absolute value ranging from 0 to 1. The correlation of 1 means that $V_\text{sd}(T)$ is perfectly linear.
\begin{equation}\label{eq:linearity}
	{\rho _\text{linear}} = \left| {\frac{{\operatorname{cov} \left( {{V_{sd}},T} \right)}}{{{\sigma _{{V_{sd}}}} \cdot {\sigma _T}}}} \right|
\end{equation}
where cov denotes the covariance, and $\sigma$ is the standard deviation. With this measure, the calibration linearity of Fig.~\ref{fig:sense_current}(a) is summarized in Table~\ref{table:Isense1}, where $I_\text{sense}=100$~mA gives the best linearity. Smaller and larger sensing currents deliver slightly worse performance.

\subsubsection{Resolution}
Apart from the linearity, an acceptable resolution matters for hardware implementation of TSEP measurements. For a given sensing current density, the temperature derivative of (\ref{eq:pn_junc}) yields
\begin{equation}
	\frac{{d{V_{pn}}}}{{dT}} =  - \frac{{{k_b}}}{q}\ln \left( {\frac{K}{j_\text{sense}}} \right)
\end{equation}
It exhibits a negative logarithmic dependency on the current density when $V_{pn}$ is the dominant factor of the device's total voltage drop. Consequently, the resolution decreases as the sensing current increases. A resolution factor $K_\text{res}$ is used to quantify the obtained calibration results and is defined as
\begin{equation}\label{eq:sensitivity}
	{K_\text{res}} = \frac{{\Delta {V_{sd}}}}{{\Delta T}}{\text{ }}[{{{\text{mV}}} \mathord{\left/
			{\vphantom {{{\text{mV}}} {\text{K}}}} \right.
			\kern-\nulldelimiterspace} {\text{K}}}]
\end{equation}
The corresponding resolution of different sensing currents are also listed in Table~\ref{table:Isense1}, where a clear drop of resolution can be seen with the increase of sensing current. For viable sampling hardware implementation, the sensitivity is recommended to be above 1~mV/K and apparently, all above scenarios can meet this requirement

\subsubsection{Self Dissipation Ratio}
Apparent self heating caused by the sensing current shall be avoided to introduce non-negligible junction temperature error. To quantify this, a self dissipation ratio ${\eta _{sd}}$ is defined as
\begin{equation}\label{eq:self_dissipation}
	{\eta _{sd}} = \frac{{{P_{sense}}}}{{{P_{rate}}}} = \frac{{{I_{sense}} \times {V_{sd@{I_{sense}}}}}}{{I_{D@{T_{ref}}}^2 \times {R_{dson}}}}
\end{equation}
where $P_\text{sense}$ is the sensing current self dissipation, and $P_\text{rate}$, $I_{D@T_{ref}}$ and $R_{dson}$ are the rated power dissipation, current and on-state resistance under a specific temperature $T_{ref}$. They all can be derived from the data sheet \cite{c2m160}. ${V_{sd@{I_{sense}}}}$ is the measured TSEP voltage as shown in Fig.~\ref{fig:sense_current}(a). Considering the worst case scenario with $V_{sd}$ being measured under 20$^\circ$C (largest voltage drop within the practical temperature range), Table~\ref{table:Isense1} illustrates the corresponding self dissipation ratios, which increase with the rising of the sensing current. Assuming a ratio less than 1\% is safe to ensure a negligible self heating, apart from 500~mA and 1000~mA, the rest of cases are acceptable in terms of self heating. 

Based on the above analysis, sensing currents under 100~mA gives the best linearity and 5~mA delivers the largest resolution and the smallest self dissipation ratio. However, the thermal transient measurement also relies on the dynamic performance, which necessitates further analysis.

\subsection{Dynamic Impact of Sensing Current Density}\label{sec:dynamic_Isense}
Electrical and thermal transients coexist during the period from \textcircled{1} to \textcircled{2} as shown in Fig.~\ref{fig:calibration} and  improper decoupling may deliver incorrect cooling curve measurement. To address this issue, standard JESD 51-1\cite{jesd511intergrated} introduces a delay time ($t_\text{MD}$) to remove the unwanted electrical transient. A further linear extrapolation method is applied to estimate the temperature at $t=0$~s. However, SiC MOSFETs is very likely to suffer long $t_\text{MD}$, which can heavily deteriorate the accuracy of the estimated temperature at $t=0$~s and lead to distortion of the obtained device thermal structure function. Thus, a shorter $t_\text{MD}$ with less electrical disturbance is needed.

Fig.~\ref{fig:sense_current}(b) shows the measured turn-off transient right after the heating phase ends. $V_\text{sd}$ rises from around -3~V (heating) to the TSEP voltage range, approximately 2.3$\sim$2.9~V in Fig.~\ref{fig:sense_current}(a). The transient duration is strongly correlated to the sensing current. With $I_\text{sense}=5$~mA, the measurement delay time is more than 600~$\mu$s, which is much longer than the time scale of chip thermal transient. If the same method in JESD 51-1 is applied, it will be unlikely to get the structure information of the chip. By increasing $I_\text{sense}$ to 100~mA, the time delay reduces to an acceptable 42~$\mu$s. Further boosting of the sensing current has a limited effect on reducing the $t_\text{MD}$ but rapidly increasing the self-dissipation ratio. Considering both the static and dynamic performances, a 100~mA sensing current achieves a better overall performance for the device under test.

\subsection{Impact of Gate Voltage on the Calibration}\label{sec:static_gate_voltage}
\begin{figure*}[!t]
	\centering
	\includegraphics[width=168mm]{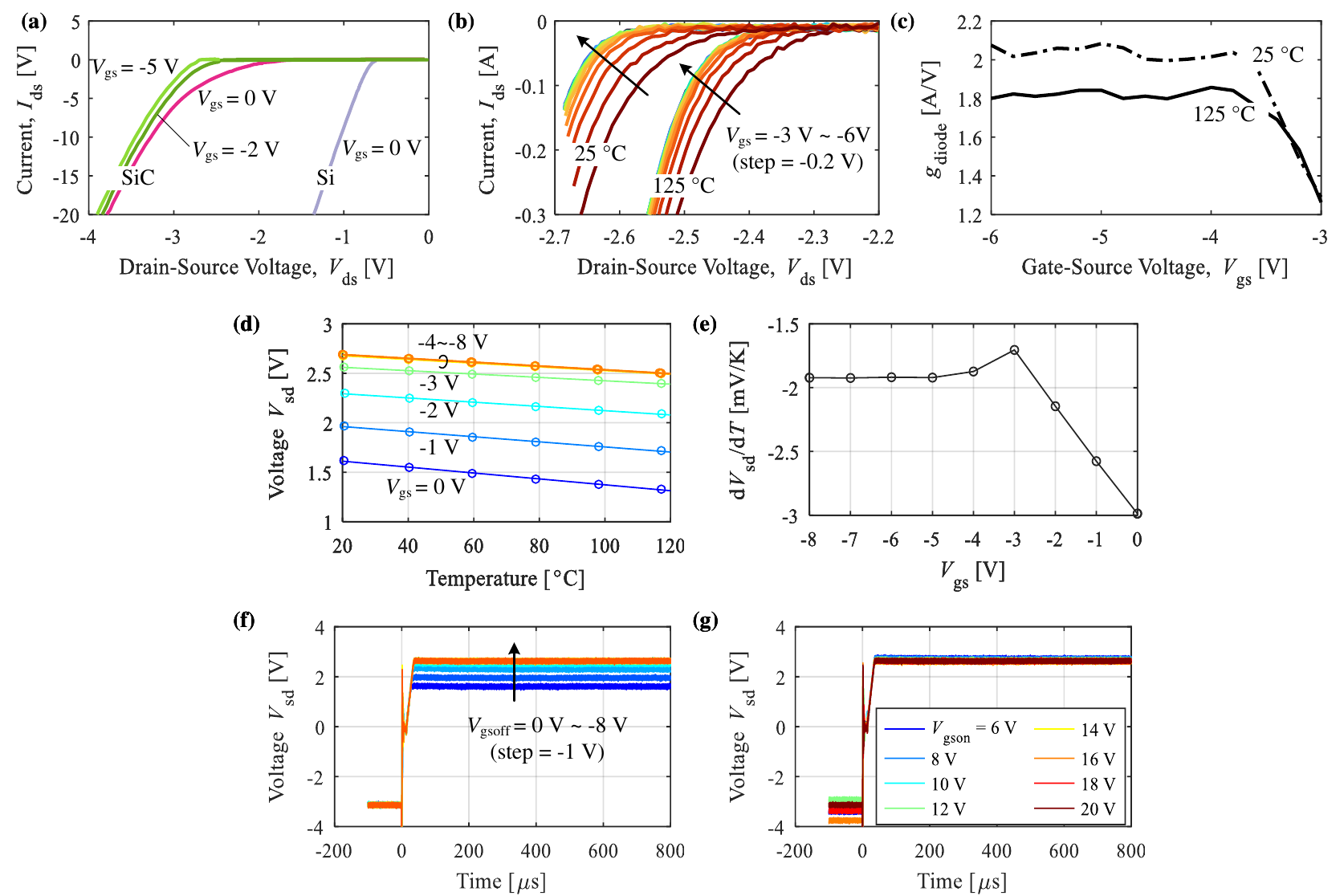}
	\caption{Static and dynamic impacts of different gate voltages: (a) data sheet comparison Si vs. SiC at 25~$^\circ$C, (b) measured output characteristics of SiC MOSFET body diode around the sensing current, (c) the proposed electrical conductance of the body diode at $I_\text{sense}=100$~mA, (d) calibration results with different $V_\text{gsoff}$ and (e) the slope of the source-drain voltage with temperature, (f) the dynamic impacts of different $V_\text{gsoff}$ and (g) $V_\text{gson}$. (The sensing current is fixed at 100~mA for the calibrations.)}
	\label{fig:calibration_vgs}
\end{figure*}

Apart from the sensing current, a precise $V_\text{sd}$ based junction temperature measurement must ensure that all sensing current goes through the body diode. Any additional channel current alters the temperature characteristics (see Fig.~\ref{fig:sic_mosfet}).  

Comparing the body diode characteristics of the SiC and Si MOSFET as shown in Fig.~\ref{fig:calibration_vgs}(a), a gate voltage of 0~V is sufficient to close the Si MOSFET channel completely, and the body diode characteristic does not change with gate voltage. By contrast, due to the wide band-gap and high forward voltage of the body diode, a negative gate voltage has to be applied to SiC MOSFETs and shall be carefully selected. In existing studies, e.g., Herold \emph{et al.}\cite{herold2017powercycling} have experimentally shown that $V_\text{gsoff}=-6$~V is the voltage to ensure a complete shut-down of the SiC MOSFET channel. However, given different die design and manufacturing process between vendors, this gate voltage may vary significantly, which thus necessitates an approach to guide the selection a proper gate turn-off voltage. In the following section, two methods are introduced


\subsubsection{Method 1: Output Characteristics under Sensing Current}
The first method depends on the output characteristic of the body diode under the sensing current level, which is not available in a typical data sheet that is derived based on larger functional currents (see Fig.~\ref{fig:calibration_vgs}(a)). 
In case the gate turn-off voltage is insufficient, e.g., $V_\text{gs}=-3$~V, the output characteristic curves shift significantly with the gate voltage, which can be seen from Fig.~\ref{fig:calibration_vgs}(b).  When the gate voltage is approaching to -6~V, the curves starts to overlap with each other regardless of temperature. 

To further quantify this effects, an electrical conductance $g_\text{diode}$ of the body diode at the sensing current is defined in (\ref{eq:g}). When the entire current flows through the internal body diode, the conductance is independent of gate voltage and becomes a constant. Thus, by evaluating the expression of (\ref{eq:g/Vgs}), it is also possible to determine the minimum $V_\text{gs}$ ensuring a completely-off channel. 
\begin{equation}\label{eq:g}
    {g_{diode}} = {\left. {\frac{{d{I_{ds}}}}{{d{V_{ds}}}}} \right|_{{I_{ds}} = {I_{sense}}}}
\end{equation}
\begin{equation}\label{eq:g/Vgs}
    \frac{{\partial {g_{diode}}}}{{\partial {V_{gs}}}} = 0
\end{equation}
Correspondingly, Fig.~\ref{fig:calibration_vgs}(c) shows the conductance of the SiC MOSFET against different gate voltages. The conductance $g_\text{diode}$ keeps stable when $V_\text{gs}<-4.5$~V.

\subsubsection{Method 2: Calibration Curves with Varied Gate Voltages}
In contrast to the output characteristics, the TSEP measuring circuit is designed for small sensing current and has a higher resolution. The calibration results with various gate voltages are shown in Fig.~\ref{fig:calibration_vgs}(d) and are upward shifted when the gate turn-off voltage drops from 0~V to -4~V. Once $V_\text{gs}$ is smaller than -4~V, the calibration curves overlap with each other. Similarly, when the MOSFET channel is fully turned off, the voltage across the pn-junction at a constant sensing current is linearly dependent on temperature. Thus, the following criterion is used according to\cite{gothner2018challenges}
\begin{equation}
	\frac{\partial }{{\partial {V_{gs}}}}\left[ {\frac{{\partial {V_{sd}}\left( T \right)}}{{\partial T}}} \right] = 0
\end{equation}
The evaluated result is shown in Fig.~\ref{fig:calibration_vgs}(e). The derivative of $V_\text{sd}(T)$ is plotted as a function of $V_\text{gs}$ and it becomes constant for $V_\text{gs}<-5$~V, which is the minimum negative voltage to ensure the channel is completely shut. Comparing the two methods, similar results of $V_\text{gs}<-5$~V and $V_\text{gs}<-4.5$~V are delivered. However, the thermal calibration needs much longer time (e.g., hours) compared with output characteristics measurement (e.g., minutes)

\subsection{Static and Dynamic Impacts of Gate Voltages}\label{sec:3d}
To evaluate the calibration results of Fig.~\ref{fig:calibration_vgs}(d), the linearity, resolution, and self dissipation ratio are also estimated according to (\ref{eq:linearity}), (\ref{eq:sensitivity}), and (\ref{eq:self_dissipation}). The corresponding results are listed in Table~\ref{table:Isense1}. When the gate voltage changes from 0~V to -3~V, the linearity is clearly deteriorated compared to the rest. The poor linearity indicates that the measured $V_\text{sd}$ is not dominated by the pn junction. Meanwhile, by adjusting the gate turn-off voltage from 0~V to -8~V, the resolution becomes weaker moderately and the self-dissipation ratio increases slightly.

The dynamic states of different turn-on or turn-off gate voltages are shown in Figs.~\ref{fig:calibration_vgs}(f) and (g). The effect of the gate voltage on the measurement delay time is almost negligible. Within the device's maximum allowable gate voltage range, a larger value brings benefits in statics and almost no side effects in the dynamics (e.g., $t_\text{MD}$).

\section{Cooling Curve Measurement and Device Structure function}
\begin{figure*}[!t]
	\centering
	\includegraphics[width=170mm]{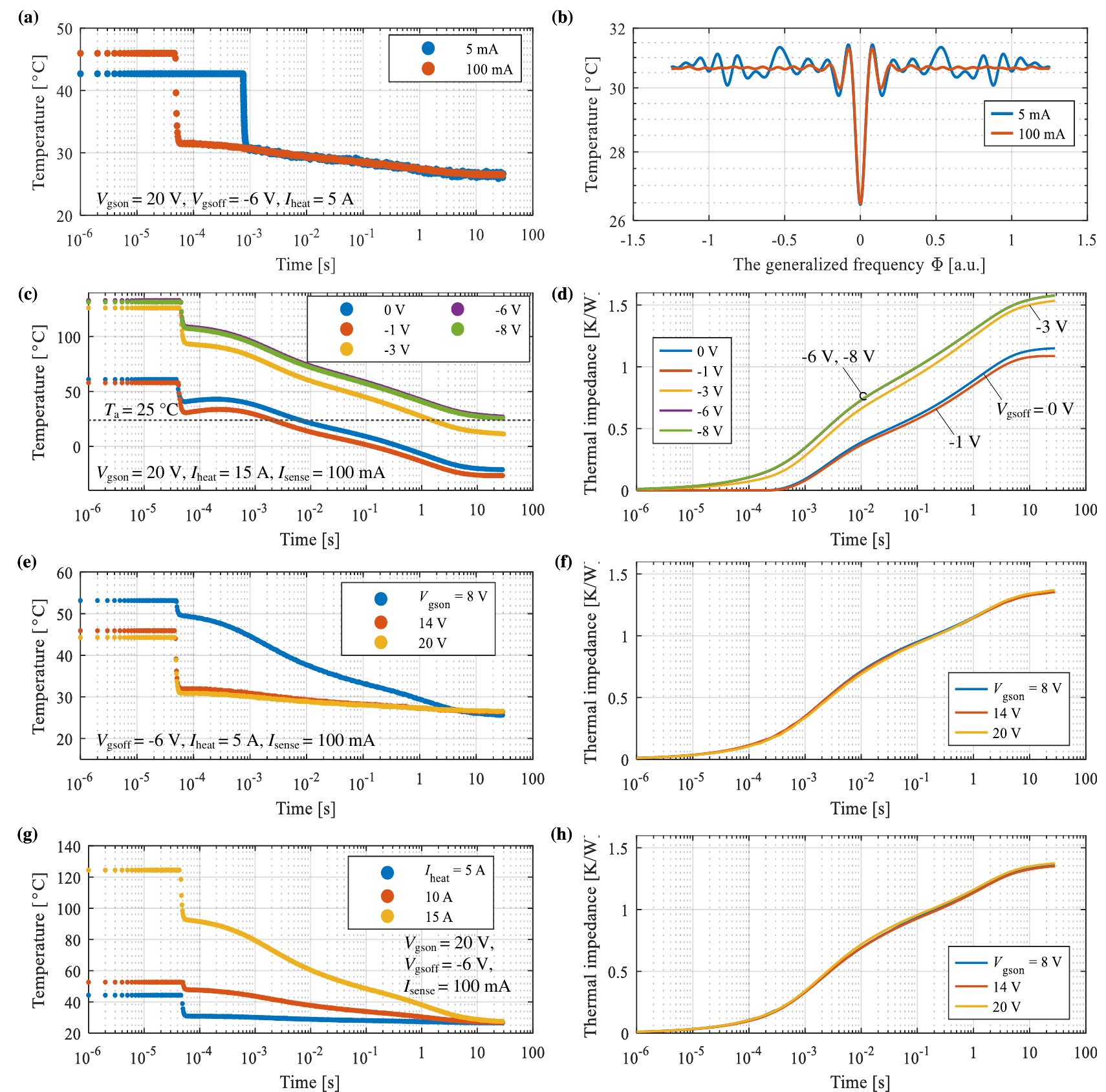}
	\caption{Cooling curve measurement with varied conditions: (a) the cooling curves under two sensing currents and (b) the corresponding frequency analysis of the measured results, (c) the cooling curves under different $V_\text{gsoff}$ and (d) the corresponding thermal impedances, (e) the cooling curves under varied $V_\text{gson}$ and (f) their thermal impedances, (g) and (h) the measurements under three different heating currents.}\vspace{-1em}
	\label{fig:varied_vgs}
\end{figure*}
\begin{figure}[!t]
	\centering
	\includegraphics[width=80mm]{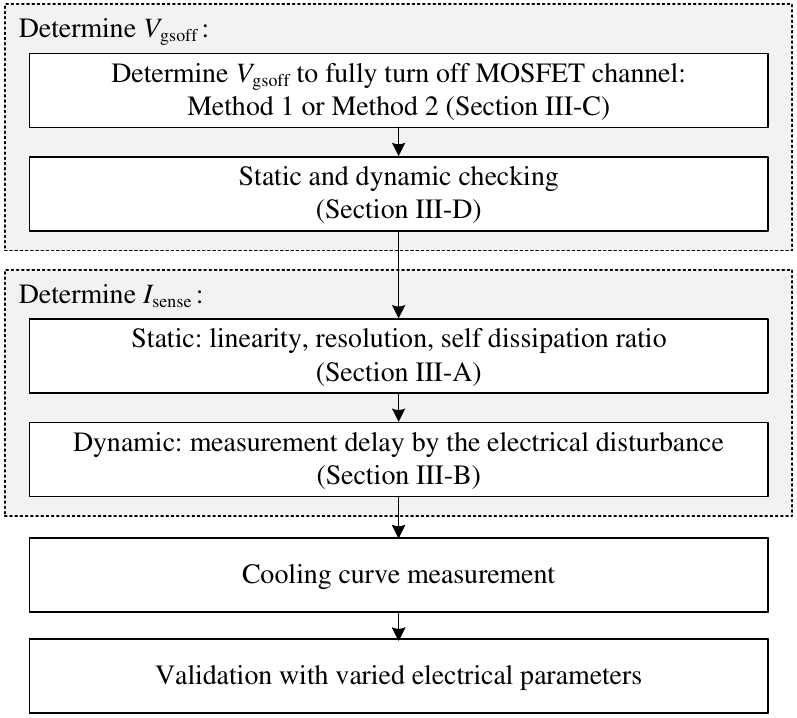}
	\caption{A flowchart to perform a reproducible transient thermal measurement for SiC MOSFETs.}
	\label{fig:flow}
\end{figure}

Once the calibration is completed, the derived relationship between $V_\text{sd}$ and temperature can be used for the cooling curve measurement. In this process, the selection of the heating current, gate turn-on voltage together with the aforementioned parameters in the calibration will be evaluated together.


\subsection{Cooling Curve Measurements and the Impacts of Key Parameters}

\subsubsection{Impact of Sensing Current}
Fig.~\ref{fig:varied_vgs}(a) shows the cooling curves of the SiC MOSFET under two different sensing currents. Ideally, the two measurements shall overlap completely. However, as it can be seen from Fig.~\ref{fig:varied_vgs}(a), the two curves differentiate until 1~ms. Specifically, for $I_\text{sense}$ = 5~mA, the impact of electrical transient is only negligible after 1 ms, which is less than 100~$\mu$s for $I_\text{sense}=100$~mA. The reason is that the turn-on of the body diode needs to build up enough minority carrier charge, which takes a longer time given a smaller sensing current. These measured results verify the aforementioned dynamic study in $\mathsection$\ref{sec:dynamic_Isense}. 

Meanwhile, it is easy to find that the measured result with $I_\text{sense}=5$~mA has more noises in Fig.~\ref{fig:varied_vgs}(a). A further frequency analysis for the duration from 1~ms to 30~s is conducted as shown in Fig.~\ref{fig:varied_vgs}(b). The high-frequency variations decay quickly for the case with $I_\text{sense}=100$~mA while the reduced sensing current has more high-frequency noise. It reveals that the electrical disturbance does not decay to zero even after $t_\text{MD}$. A continuous effect exists within the entire measurement although it is not as dominant as in the first phase.

\subsubsection{Impact of Gate Turn-Off Voltage}
$\mathsection$\ref{sec:static_gate_voltage} emphasizes that an insufficient gate turn-off voltage develops a parallel current path via the MOSFET channel, which deteriorates the temperature measurement. However, Fig.~\ref{fig:calibration_vgs}(d) shows that a good calibration result still can be obtained regardless of the value of the gate voltage, but their correspondingly measured thermal impedances are different as shown in Fig.~\ref{fig:varied_vgs}(c). When the gate turn-off voltage is severely insufficient (e.g., 0~V and -1~V), the measured junction temperature rises at approximately 2$\times 10^{-4}$~s after shutting off the heating current. As the measured cooling stage does not have heat injection, the estimated junction temperature based on the calibration result is inconsistent with the physical phenomenon. This behavior has also been observed in\cite{funaki2016difficulties} with a conclusion of imperfect structure of the SiC MOSFET. Based on the investigation in this paper, an additional reason is the insufficient gate voltage, which can, at the meantime, lead to incorrect measurements that are below the ambient temperature 25$^\circ$C. With a lower gate turn-off voltage, e.g., $V_\text{gsoff}=-3$~V, the temperature rises after the heating current shuts off becomes invisible. However, the final temperature at 30~s is still below 25$^\circ$C. These inconsistencies reveals the significance of the gate turn-off voltage. By further reducing gate voltage to -6~V and -8~V, the measured cooling results are consistent regardless of the gate voltage.

Thermal impedances are shown in Fig.~\ref{fig:varied_vgs}(d). In essence, the thermal impedance reflects the thermal structure of a semiconductor package, which should not vary with the gate voltage. However, when the gate turn-off voltage is insufficient, the obtained thermal impedance curves are underestimated. By contrast, a stable thermal impedance curve can be obtained under $V_\text{gsoff}\le-6$~V.

\subsubsection{Impact of Gate Turn-On Voltage}
Gate turn-on voltage ($V_\text{gs\_on}$) decides the channel voltage drop in the heating stage. As shown in Fig.~\ref{fig:varied_vgs}(e), with different gate turn-on voltages but an identical heating current, the maximum temperature difference is up to 20$^\circ$C. However, the derived thermal impedance curves barely change as shown in Fig. \ref{fig:varied_vgs}(f). Besides, the measurement delay time does not change with $V_\text{gs\_on}$. Thus, $V_\text{gs\_on}$ does not affect the thermal characterization.

\subsubsection{Impact of Heating Current}
The heating current affects the power dissipation to heat the device up. As shown in Fig.~\ref{fig:varied_vgs}(g), a higher heating current leads to a higher junction temperature. However, the thermal impedance curves completely overlap as shown in Fig.~\ref{fig:varied_vgs}(h). Thus, $I_\text{heat}$ does not significantly affect the thermal impedance characterization.

In summary, the calibration parameters in $\mathsection$\ref{sec:calibration} are critical for achieving correct cooling curve measurement. The other two parameters (i.e., $V_\text{gson}$ and $I_\text{heat}$) have a negligible effect. In addition, the consistency of the thermal impedance curves under different electrical parameters can be used to justify if thermal impedance curve is correctly measured because thermal structure of a semiconductor package does not inherently vary with the testing electrical parameters.

\subsection{ Measurement Flow to Achieve a Reproducible Transient Thermal Measurement}
A flowchart to achieve a reproducible transient thermal measurement for SiC MOSFETs is provided as shown in Fig.~\ref{fig:flow}. The aforementioned analysis indicates that the gate turn-off voltage is critical to perform a correct measurement, and shall be determined at the very beginning. Either Method 1 or Method 2 in $\mathsection$\ref{sec:static_gate_voltage} can be applied. A certain margin can be added within the maximum allowable gate voltage because a large negative voltage benefits both the static and dynamic states. Next, a sensing current is selected carefully. Either too large or too small sensing current is not conducive to the transient thermal measurement. To ensure the pn-junction dominates the measured $V_\text{sd}$ voltage, both the static and the dynamic states in terms of the linearity, resolution, self dissipation ratio, and the measurement delay should be evaluated comprehensively. The following cooling curve measurement can be conducted with the selected testing parameters. A final validation process can be added with varied heating currents or gate turn-on voltages.

\section{Conclusion}
This paper investigates the thermal characterization method of SiC MOSFET based on the source-drain voltage of the body diode. The two key steps in terms of the calibration and the cooling curve measurement are evaluated comprehensively. In the calibration, both the static and dynamic states of the selection of the sensing current and the gate turn-off voltage are evaluated regarding their linearity, resolution, self dissipation ratio, and electrical disturbance. Two methods are proposed to select the gate turn-off voltage in order to ensure the MOSFET channel is completely turned off in temperature measurement. Based on the above analysis, we obtained following conclusions:
\begin{itemize}
	\item[1)] A large negative gate turn-off voltage benefits both the static and the dynamic transient temperature measurement, while too large or small sensing current is not ideal for the measurement of SiC MOSFETs;
	\item[2)] Insufficient negative gate turn-off voltage leads to an incorrect temperature measurement in the cooling curve, e.g., abnormal temperature rise or measured temperature below ambient conditions. These results violating physics emphasize the importance of the gate turn-off voltage.
	\item[3)] Insufficient sensing current deteriorates the dynamics in terms of longer electrical disturbance and more noises, while too large sensing current sacrifices the steady-state performance in particular of a large self dissipation ratio.
	\item[4)] The gate turn-on voltage and the heating current have negligible impacts on the measured thermal impedance. As the thermal structure of a semiconductor package does not inherently vary with these two parameters, the consistency of the thermal impedance curves under varied gate turn-on voltage or heating current can be used as a validation.
\end{itemize}
To perform a reproducible transient thermal measurement for SiC MOSFETs, a guide flowchart is provided in this paper, which includes the selection of the electrical parameters and a validation process.

\section*{Acknowledgement}
This research has been supported by Independent Research Fund Denmark with the number 1031-00024B.

\ifCLASSOPTIONcaptionsoff
  \newpage
\fi

\bibliographystyle{IEEEtran}
\bibliography{MMClifetime}

\begin{thebibliography}{10}
\providecommand{\url}[1]{#1}
\csname url@samestyle\endcsname
\providecommand{\newblock}{\relax}
\providecommand{\bibinfo}[2]{#2}
\providecommand{\BIBentrySTDinterwordspacing}{\spaceskip=0pt\relax}
\providecommand{\BIBentryALTinterwordstretchfactor}{4}
\providecommand{\BIBentryALTinterwordspacing}{\spaceskip=\fontdimen2\font plus
\BIBentryALTinterwordstretchfactor\fontdimen3\font minus
  \fontdimen4\font\relax}
\providecommand{\BIBforeignlanguage}[2]{{%
\expandafter\ifx\csname l@#1\endcsname\relax
\typeout{** WARNING: IEEEtran.bst: No hyphenation pattern has been}%
\typeout{** loaded for the language `#1'. Using the pattern for}%
\typeout{** the default language instead.}%
\else
\language=\csname l@#1\endcsname
\fi
#2}}
\providecommand{\BIBdecl}{\relax}
\BIBdecl

\bibitem{jesd511intergrated}
JEDEC, ``{EIA/JESD} 51-1: Intergrated circuits thermal measurement
  method-electrical test method (single semiconductor device),'' 1995.

\bibitem{JEDEC51-14}
------, ``{JESD51-14: Transient dual interface test method for the measurement
  of the thermal resistance junction-to-case of semiconductor devices with heat
  flow through a single path},'' 2010.

\bibitem{gao2020two}
S.~Gao, K.~D. Ngo, and G.-Q. Lu, ``Two-dimensional mapping of interface thermal
  resistance by transient thermal measurement,'' \emph{IEEE Trans. Ind.
  Electron.}, vol.~68, no.~5, pp. 4448--4456, 2020.

\bibitem{schweitzer2008transient}
D.~Schweitzer, H.~Pape, and L.~Chen, ``Transient measurement of the
  junction-to-case thermal resistance using structure functions: chances and
  limits,'' in \emph{Proc. 24th Annual IEEE Semicond. Therm. Meas. Manage.
  Symp.}, 2008, pp. 191--197.

\bibitem{kestler2018junction}
T.~Kestler and M.~Bakran, ``Junction temperature measurement of {SiC MOSFETs}:
  Straightforward as it seems,'' in \emph{Proc. PCIM}, 2018, pp. 1--6.

\bibitem{yangfei2019design}
F.~Yang, E.~Ugur, and B.~Akin, ``{Design methodology of DC power cycling test
  setup for SiC MOSFETs},'' \emph{IEEE J. Emerging Sel. Top. Power Electron.},
  vol.~8, no.~4, pp. 4144--4159, 2019.

\bibitem{bhatnagar1993comparison}
M.~Bhatnagar and B.~J. Baliga, ``{Comparison of 6H-SiC, 3C-SiC, and Si for
  power devices},'' \emph{IEEE Trans. Electron Devices}, vol.~40, no.~3, pp.
  645--655, 1993.

\bibitem{gonzalez2017impact}
J.~O. Gonzalez and O.~Alatise, ``{Impact of the gate driver voltage on
  temperature sensitive electrical parameters for condition monitoring of SiC
  power MOSFETs},'' \emph{Microelectron. Reliab.}, vol.~76, pp. 470--474, 2017.

\bibitem{funaki2016difficulties}
T.~Funaki and S.~Fukunaga, ``{Difficulties in characterizing transient thermal
  resistance of SiC MOSFETs},'' in \emph{Proc. 22nd Int. Workshop Therm.
  Invest. ICs Systems (THERMINIC)}.\hskip 1em plus 0.5em minus 0.4em\relax
  IEEE, 2016, pp. 141--146.

\bibitem{zeng2018difference}
G.~Zeng, H.~Cao, W.~Chen, and J.~Lutz, ``Difference in device temperature
  determination using pn-junction forward voltage and gate threshold voltage,''
  \emph{IEEE Trans. Power Electron.}, vol.~34, no.~3, pp. 2781--2793, 2018.

\bibitem{Lutz2011Semi_book}
J.~Lutz, H.~Schlangenotto, U.~Scheuermann, and R.~D. Doncker,
  \emph{{Semiconductor Power Devices: Physics, Characteristics,
  Reliability}}.\hskip 1em plus 0.5em minus 0.4em\relax Springer, 2011.

\bibitem{c2m160}
``{CREE C2M0160120D} silicon carbide power {MOSFET},''
  \url{https://assets.wolfspeed.com/uploads/2020/12/C2M0160120D.pdf},
  access:2022-08-05.

\bibitem{herold2017powercycling}
C.~Herold, J.~Sun, P.~Seidel, L.~Tinschert, and J.~Lutz, ``Power cycling
  methods for sic mosfets,'' in \emph{Proc. 29th Int. Symp. Power Semicond.
  Devices \& IC's (ISPSD)}.\hskip 1em plus 0.5em minus 0.4em\relax IEEE, 2017,
  pp. 367--370.

\bibitem{gothner2018challenges}
F.~Gothner, O.~C. Spro, M.~Herncs, and D.~Peftitsis, ``{Challenges of SiC
  MOSFET power cycling methodology},'' in \emph{Proc. 20th Eur. Conf. Power
  Electron. Appl. (EPE'18 ECCE Europe)}.\hskip 1em plus 0.5em minus 0.4em\relax
  IEEE, 2018, pp. P--1.

\end{thebibliography}

\end{document}